# Interface-driven unusual anomalous Hall effect in Mn$_x$Ga/Pt bilayers: No correlation with chiral spin structures

Kangkang Meng [1, *], Lijun Zhu [2], Zhenhu Jin [3], Enke Liu [4], Xupeng Zhao [5], Iftikhar Ahmed Malik [6], Zhenguo Fu [7], Yong Wu [1], Jun Miao [1], Xiaoguang Xu [1], Jinxing Zhang [6], Jianhua Zhao [5] and Yong Jiang [1, *]

[1] *Beijing Advanced Innovation Center for Materials Genome Engineering, School of Materials Science and Engineering, University of Science and Technology Beijing, Beijing, China*

[2] *Cornell University, Ithaca, New York 14853, USA*

[3] *Department of Applied Physics, Graduate School of Engineering, Tohoku University, 6-6-05, Aoba-yama, 980-8579 Sendai, Japan*

[4] *Institute of Physics, Chinese Academy of Sciences, Beijing, China*

[5] *State Key Laboratory of Superlattices and Microstructures, Institute of Semiconductors, Chinese Academy of Sciences, Beijing, China*

[6] *Department of Physics, Beijing Normal University, Beijing, China*

[7] *Institute of Applied Physics and Computational Mathematics, Beijing, China*

**Abstract:** The effects of spin-orbit coupling and symmetry breaking at the interface between a ferromagnet and heavy metal are particularly important for spin-based information storage and computation. Recent discoveries suggest they can create chiral spin structures (e.g. skyrmions), which have often been identified through the appearance of the bump/dip features of Hall signals, the so-called topological Hall effect (THE). In this work, however, we have present an unusual anomalous Hall effect (UAHE) in Mn$_x$Ga/Pt bilayers and demonstrated that the features extremely similar to THE can be generated without involving any chiral spin structures. The low temperature magnetic force microscopy has been used to explore the magnetic field-dependent behavior of spin structures, and the UAHE as a function of magnetic field does not peak near the maximal density of magnetic bubbles. The results unambiguously evidence that the UAHE in Mn$_x$Ga/Pt bilayers shows no correlation with chiral spin structures but is driven by the modified interfacial properties. The bump/dip features of Hall signals cannot be taken as an unambiguous signature for the emergence of chiral spin structures, and a wealth of underlying and interesting physics need explored.

*Corresponding author:

*kkmeng@ustb.edu.cn

*yjiang@ustb.edu.cn

Interface-driven magnetic effects and phenomena associated with spin-orbit coupling (SOC) and intrinsic symmetry breaking at interfaces are at the forefront of condensed-matter physics research [1]. Effects at interfaces can be classified as emergent in the sense that complex unanticipated phenomena emerge from apparently simple materials and interactions, providing challenges of predictability and design of functionality [2, 3]. It has only recently been appreciated that interfaces, particularly those where SOC is strong, can fundamentally change the magnetic ground states in the ferromagnet (FM)/heavy metal (HM) heterostructures. Owing to the presence of strong SOC and inversion asymmetry, Dzyaloshinskii-Moriya interaction (DMI), an antisymmetric exchange interaction that favors a chiral arrangement of the magnetization, will be generated [2-4]. It is in the form of $\boldsymbol{D}_{12} \cdot (\boldsymbol{S}_1 \times \boldsymbol{S}_2)$, where the vector $\boldsymbol{D}_{12}$ represents the DMI strength, $\boldsymbol{S}_1$ and $\boldsymbol{S}_2$ are the total spin of two nearby atoms [5, 6]. When the DMI is sufficiently strong compared to other interactions, it could lead to topologically protected (chiral) spin structures, such as skyrmions or skyrmion lattices [7]. Skyrmions can be defined by the topological number $N$, which is a scale of the winding of the normalized local magnetization, $\boldsymbol{m}$. In the two-dimensional limit, the topological number is $N = \frac{1}{4\pi} \int \boldsymbol{m} \cdot (\partial_x \boldsymbol{m} \times \partial_y \boldsymbol{m}) \mathrm{d}x\mathrm{d}y$ [3, 7]. This non-trivial topological property governs some of the most important properties of skyrmions including the topological Hall effect (THE) [8-10]. When a conduction electron passes through a skyrmion, the spin of the conduction electron will adiabatically follow the texture and acquire a real-space Berry phase, which deflects the conduction electrons perpendicular to the current direction and causes an additional contribution to the observed Hall signals termed as THE, characterized by the appearance of bumps or dips in the Hall measurements. The effective magnetic field can be described by $B_{\mathrm{eff}} = \frac{\hbar c}{2e} \hat{z} \boldsymbol{m} \cdot (\partial_x \boldsymbol{m} \times \partial_y \boldsymbol{m})$, which is closely related to the definition of topological number (3). However, the above real-space Berry phase picture is valid only if the exchange coupling between electrons and local magnetization is strong (adiabatic approximation), while it fails in the weak coupling regime (non-adiabatic approximation) since the electrons fail to adjust their spin to the local magnetization and the spin-flip processes are activated [11-13]. In such a regime, the electrons will experience an inhomogeneous emergent field as there is randomly distributed chiral spin structures, in contrast to electrons in bulk systems which experience a systematically varying uniform emergent magnetic field due to the ordered arrangement of chiral spin structures such as skyrmion crystals [14].

In this work, however, we report an unusual anomalous Hall effect (UAHE) that can generate the same transport features of the THE without involving any chiral spin structures. Low temperature magnetic force microscopy (MFM) results indicate that the additional Hall signals as a function of magnetic field at 3.5 K does not peak near the maximal density of magnetic bubbles. The results unambiguously evidence that the UAHE in $Mn_x$Ga/Pt bilayers shows no correlation with chiral spin structures but is driven by the modified interfacial properties due to the emergence of strong interfacial SOC introduced by Pt. The bump/dip features of Hall signals cannot be taken as an unambiguous signature for the emergence of chiral spin structures, and it should be treated discreetly since a wealth of underlying and interesting physics are often regrettably missed.

For this study, we prepared magnetic bilayers of $Mn_x$Ga 6 nm/Pt 5 nm ($x$=1.0, 1.25, 1.55) and a reference sample of $Mn_x$Ga 6 nm/Al 5 nm (see also Materials and Methods). The 1-nm-thick $Mn_x$Ga films were firstly grown on 100-nm-thick GaAs buffered semi-insulating GaAs (001) substrates at 70℃ by molecular-beam epitaxy with controlling the flux of Mn and Ga atoms. Then the films were annealed at 300 ℃ for 1 min, and other 5-nm-thick MnGa films were continued to be grown at 300 ℃ after the annealing. Finally, the films were transferred to the magnetron sputtering system immediately through a lower vacuum chamber, and 5-nm-thick Al or Pt films were deposited on top of the $Mn_x$Ga films. All the films have been patterned into Hall bars in the size of 10 μm × 80 μm using photolithography and Ar ion milling. Similar to our previous work [15-17], both the Hall resistivity and longitudinal resistivity in this work were determined by subtracting the contributions of the $Mn_x$Ga single layer by assuming a parallel resistor model. By subtracting the $H$-linear ordinary Hall term ($\rho_0 = R_0 H$) from the total Hall resistivity $\rho_{XY}$, we have obtained the sum of other Hall contributions of the $Mn_xGa/Pt$ bilayers under perpendicularly applied magnetic field ($H$) as shown in Figure 1 (a)-(c). Bumps or dips emerge in the whole temperature range from 5 to 300 K for all the three bilayers. Fig. 1 (d)-(f) show the temperature dependence of magnetoresistance (MR), $MR = (\rho_{XX}(H) - \rho_{XX}(0))/\rho_{XX}(0)$, under perpendicular magnetic field. The sign of MR changes from negative to positive as the temperature decreases. Both the Hall and longitudinal signals are quite different from those of $Mn_xGa/Al$ bilayers [18], in which the transport properties are determined by the single $Mn_xGa$ layers [16]. Compared to $Mn_xGa/Al$, the $Mn_xGa/Pt$ bilayers show distinct temperature dependences of anomalous Hall resistivity $\rho_A$ and longitudinal resistivity $\rho_{XX}$ [18]. It is found that $\rho_A$ can be described by scaling law $\rho_A = \alpha\rho_{XX0} + b\rho^2_{XX}$ with $\rho_{XX}$ for the $Mn_xGa/Al$ bilayers [18], where $\alpha$ and $b$ denote extrinsic skew scattering and intrinsic mechanism respectively, $\rho_{XX0}$ is the residual resistivity induced by impurity scattering [19]. We believe that 6 nm $Mn_xGa$ is thick enough to prevent the deterioration of its bulk magnetic properties, and new magnetic states or transport features are expected to be of interfacial nature. Then, we attempt to investigate the anomalous Hall effect (AHE) in the $Mn_xGa/Pt$ bilayers using the same scaling law. Fig. 2 (a)-(c) show the experimental and fitted relationships between $\rho_A$ and $\rho^2_{XX}$. At relatively high temperatures, the scaling law is found to work only for large $\rho_{XX}$. The slope $b$ of the $Mn_xGa/Pt$ bilayers is similar to that in the $Mn_xGa/Al$ bilayers with the same Mn component $x$, indicating the same intrinsic nature of the $Mn_xGa$ films. In the low temperature limit, there are large positive deviations, which should be ascribed to the increase of $\rho_A$ since $\rho_{XX}$ decreases monotonically with decreasing temperature and then approaches a constant due to weak Kondo effect [18]. The upturn of $\rho_A$ is an indicative of the onset of large extrinsic skew scattering due to the strong SOC of impurities. The temperature dependence of $\alpha$ has been shown in Fig. 2(d), and a deviation has been found in the low temperature limit.

After subtracting ordinary and anomalous Hall signals, we can get additional Hall resistivities $\Delta\rho_{XY} = \rho_{XY} - \rho_0 - \rho_A$ as shown in Figs. 3(a)-(c), and the shapes of $\Delta\rho_{XY}(H)$ reverse in the low temperature limit with sweeping magnetic field from positive to negative. The temperature dependences of the largest $\Delta\rho_{XY}$ ($\Delta\rho^{max}_{XY}$) in all the bilayers have been shown in Fig. 3(d), and a sign change is observed for all the bilayers at low temperature. Here, we use MFM to determine the correspondence of transport features with the spin structures. Fig. 3(e) shows an illustrative example of such images in the $Mn_{1.55}Ga/Al$ bilayers after saturating with a positive magnetic field of $H$=5 to 0 T at 300 K. The magnetization is almost fully aligned in the positive magnetic field direction and the MFM image shows several magnetic bubbles. For the $Mn_{1.55}Ga/Pt$ bilayers as shown in Fig. 3(f), a dilute density of small magnetic bubbles in the size of 100~200 nm have emerged among the worm-like or the long stripes patterns. It suggests that the magnetic bubbles have emerged in both the $Mn_{1.55}Ga/Al$ and $Mn_{1.55}Ga/Pt$ bilayers, while only the latter exhibits non-zero $\Delta\rho_{XY}$ at the zero magnetic field. The most different mechanisms in these two bilayers is that a strong interfacial SOC is only introduced in $Mn_{1.55}Ga/Pt$ bilayers, which will profoundly modify both the magnetic and transport features. In the presence of interfacial DMI induced by strong SOC, the domain wall configuration in the $Mn_{1.55}Ga/Pt$ bilayers may be in Néel type with fixed chirality [20]. On the contrary, in the $Mn_{1.55}Ga/Al$ bilayers that lack interfacial DMI, the magnetic bubbles were solely stabilized by strong dipole interactions that give rise to a rich collection of spin structures with non-uniform spin chirality [21]. Therefore, the chirality of the spin structures in the $Mn_xGa/Pt$ bilayers cannot be identified, and we first speculate that the non-zero $\Delta\rho_{XY}(0)$ at 300 K in $Mn_{1.55}Ga/Pt$ bilayers is the THE.

Before the discussion of the possible THE, we should elucidate whether the non-zero $\Delta\rho_{XY}$ in the $Mn_xGa/Pt$ bilayers was described in the strong (adiabatic approximation) or weak coupling (non-adiabatic approximation) regime. In this work, we use the model developed by Nakazawa $et\ al.$ [12, 22], which considers three regimes depending on the strength of the exchange coupling. The three regimes correspond to different conditions between several important system parameters, such as the characteristic length scale of the chiral spin structures $q^{-1}$, the electron mean free path $l$, the momentum relaxation time $\tau = \dfrac{m^*}{e^2\rho_{XX}n} = \dfrac{m^*R_0}{e\rho_{XX}}$, the exchange time $\tau_{ex} = \dfrac{\hbar}{2J}$. Here, $m^*$ is the effective electron mass, $R_0$ is the ordinary Hall coefficient [18], $J$ is the exchange coupling constant. The spin-resolved band structures of $L1_0$-MnGa without SOC have been calculated using first-principle calculations [18], and we can get $E_F = 5.2712$ eV, $m^* = 0.65m_0$, $J = 1.1$ eV, $k_F = 0.121\text{nm}^{-1}$. Then, we can get the corresponding constant of the $Mn_{1.55}Ga/Pt$ bilayers at 300 K, $l = \dfrac{6\pi^2\hbar}{e^2\rho_{XX}k_F^2} = 1.1\text{nm}$, $\tau = 6\times10^{-11}\text{s}$, $\tau_{ex} = 3.8\times10^{-16}\text{s}$. It is found that the adiabatic approximation is applicable since $\tau_{ex} \ll \tau$ and $(ql)^2 < 1$, indicating the spin of the conduction electrons can adiabatically follow the surrounding local magnetization. Therefore, the additional Hall signals can be written as $\Delta\rho_{XY} = PR_0B_{\text{eff}} = PR_0n_\phi^T\phi_0$, where $P$ is the spin polarization of charge carriers and determined to be 20% according to the band structures, $B_{\text{eff}}$ the fictitious magnetic field, $\phi_0 = h/e$ the flux quantum, and $n_\phi^T$ the density of chiral spin structures [2]. Then the $n_\phi^T$ of the $Mn_{1.55}Ga/Pt$ bilayers at 300 K is calculated to be 75 $\mu m^{-2}$, the separation of the chiral spin structures is ($(n_\phi^T)^{-1/2}$) is ~115 nm. Obviously, it reveals large discrepancy as compared with the MFM results, indicating the AHE in $Mn_xGa/Pt$ bilayers should be much more involved.

Large discrepancy between MFM and transport features has also been found at low temperatures. The magnetic field-dependent MFM, $M-H$ and $\Delta\rho_{XY}-H$ curves of the $Mn_{1.55}Ga/Pt$ bilayers at 3.5 K have been compared as shown in Fig. 4 to gain more insight into the nature of the spin structures and transport features. The magnetization is almost fully aligned in the positive magnetic field direction with $H=5$ T and $\Delta\rho_{XY}$ is suppressed as shown in Fig. 4 (a) and (b) respectively, and the MFM image shows a homogeneous red contrast as shown in Fig. 4(c). The $\Delta\rho_{XY}$ firstly decreased from zero to negative values with sweeping the magnetic field from positive. Although the MFM image is almost the same between 5 T and 2 T, the value of $\Delta\rho_{XY}$ is obviously non-zero at 2 T. At remanent state, the MFM image shows obscure worm-like configurations in the so-called labyrinthine state. As $H$ decreases further towards large negative values -0.5 T, the worm-like or the long stripes patterns become clear, among which several bubble-like magnetic domains have emerged. With further decreasing the magnetic field to -1.0 T that is at around the coercivity, many isolated bubble-like magnetic domains in a typical size of ~50-200 nm have been observed, while the value of $\Delta\rho_{XY}$ has approached to be zero. By the way, the same MFM measurements have also been carried out in the $Mn_{1.55}Ga/Al$ bilayers at 3.5 K [18]. A similar magnetic field-dependent MFM result has also been found in the $Mn_{1.55}Ga/Al$ bilayers, while there are no bump/dip features of Hall signals [18]. It reveals that the magnetic bubbles, in the form of up-magnetized domains in a down-magnetized background of a perpendicular magnetic anisotropy $Mn_xGa$ films, are considered to be stabilized by magnetic dipole interactions, which should be achiral. Therefore, the THE which is the hallmark of chiral spin structures is unlikely to be the origin of the bump/dip features of Hall signals in the $Mn_xGa/Pt$ bilayers. Although the THE signal arises from the chiral spin structures in magnetic multilayers, we note that the appearance of the bump/dip features of Hall signals alone is not an unambiguous evidence. In the following discussion, the additional Hall signals in the $Mn_xGa/Pt$ bilayers have been termed as UAHE, which should be strongly related to the strong interfacial SOC.

To further investigate the modified interfacial properties, the Mn $2p$ x-ray-absorption spectroscopy (XAS) spectra were recorded at room temperatures in the Total Electron Yield (TEY) mode by measuring the drain current of the electrically isolated sample as shown in Fig. 5(a). The XAS spectra of Mn atoms should mainly come from the interfaces, since it is well known that the TEY detection is characterised by a probing depth lower than 5 nm (The Al and Pt layers are all 5-nm-thick), and the number of electrons that reach the surface decays exponentially as a function of the depth of the photon absorption. In the spectra of the two bilayers, a sharp peak structure and a doublet structure are observed in the Mn $2p_{3/2}$ ($L_3$) and $2p_{1/2}$ ($L_2$) core excitation regions, respectively. In addition, some more evident shoulder structures are found on the larger energy side of the $2p_{3/2}$ component in the $Mn_{1.25}Ga/Pt$ bilayers, which is a characteristic of primarily $d^5$ with some $d^6$ ground states [27, 28]. The distinctly split structure of the $2p_{1/2}$ component has also been found, and it is similar with antiferromagnetic MnPt films, where the $3d$ orbitals of Mn are more localized [29]. It indicates a significant influence of the coordination on the electronic structure of Mn $d$ states at the interface of the $Mn_xGa/Pt$ bilayers. Though the Pt has high spin susceptibility originating from the large stoner exchange parameter, a very small magnetic moment in Pt is considered to be induced at the interfaces [30]. However, the magnetic ground state of Mn is antiferromagnetic, and a large negative nearest-neighbor exchange interaction has been theoretically confirmed by Belabbes *et al* [30]. Therefore, the interfacial ferromagnetic states of $Mn_xGa$ will be tremendously modified by the *3d-5d* orbital hybridization, leading to a combination of ferromagnetism and the antiferromagnetism at the interface. Consequently, the evolution of the electronic structure due to modulated magnetic states at the interfaces will generate significant momentum-space Berry phase contributions to the AHE at finite temperature. At high magnetic field, only the ferromagnetism will be induced and dominate the AHE. As the magnetic field decreases, the antiferromagnetism becomes increasingly evident and results in a modifed momentum-space Berry curvature at the interface. This also explains why the largest UAHE have been found at around zero magnetic field, where the antiferromagnetism may be the strongest. The phenomenon may also depend on the surface roughness of $Mn_xGa$, since the largest UAHE has been found for $x$=1.0, which also shows the largest surface roughness [18]. The increased surface roughness may decrease the directional arrangement of magentic moment of $Mn_xGa$ and assist the emergence of antiferromagnetism. Furthermore, the formation of long range skyrmions lattices at the interfaces can be prevented by large surface roughness, since it may enhance the magnetic dipole energy which will dominate the interfacial DMI and favors achiral Bloch walls [31, 32], although it has been theoretically proved that the spin-flip excitations through SOC in half-filled or high spin *3d* overlayer (Mn) make the largest contribution to DMI [30]. For further physical insight, we shows the temperature dependence of the largest additional Hall conductivity $\Delta\sigma_{XY}^{max}$ which is defined as $\Delta\rho_{XY}^{max}/\rho_{XX}^2$, as shown in Fig. 5(b). Phenomenological reasoning suggests that the longitudinal conductance increases with decreasing temperature while $\Delta\sigma_{XY}^{max}$ remains almost constant at high temperature regime which should stem from the intrinsic contributions (momentum-space Berry curvatures). In the low temperature limit, $\Delta\sigma_{XY}^{max}$ reveals a sharp decrease to large negative values. We have plotted the temperature dependence of $d^2\Delta\sigma_{XY}^{max}/dT^2$ and $d^2\alpha/dT^2$ as shown in Fig. 5(c), which show sharp variations at almost the same temperature (50 K). Notably, the aforementioned skew scattering contribution dominates the AHE in the low temperature limit as shown in Fig. 2. Therefore, the sign change of UAHE should be related to the skew scattering due to the emergence of strong interfacial SOC, which depends on the distribution, the type and the density of impurities [30-32]. Since the sign of the skew scattering contribution changes when that of the scattering potential is reversed, the impurity potential is considered to be modified through the emergence of antiferromagnetic states at around zero magnetic fields. Further theoretical work is necessary to accurately describe this complex correlation of UAHE and $3d$-$5d$ orbital hybridization in the FM/HM bilayers.

Finally, we want to mention another role of the strong interfacial SOC introduced by Pt. It is known that, in materials with heavy elements, the band inversion can be induced since the strong SOC can split the $p$ band by a large enough magnitude to flip the $s$-$p$ band structure, leading to a topological insulator [33]. The inverted bulk band structure will topologically give rise to metallic surface states, which have a nearly linear energy-momentum relationship. Interestingly, the topological surface states also exist on the surface of Weyl metals, in which the bulk bands are gapped by SOC in the momentum space except at Weyl or Dirac nodes. In the case of the Mn$_x$Ga/Pt bilayers, we also suppose the emergence of Weyl metallic features at the interfaces, which may explain the positive MR in the low temperature limit as shown in Fig. 1. The chiral anomaly is predicted to occur in Weyl metals since the conservation of chiral charges is violated in the case of a parallel magnetic and electric field, resulting in a negative longitudinal MR [34, 35]. It suggest that there is a current along the magnetic field arising from the unequal occupation of left and right moving chiral modes, giving a current of $j_c \propto B\mathbf{E}\cdot\mathbf{B}$, where $\mathbf{E}$ and $\mathbf{B}$ is the external electric and magnetic field respectively [34, 35]. Then the electric currents measured in a thin film with chiral anomaly inevitably include this additional contribution. In the case of $\mathbf{E}/\!/\mathbf{B}$, the total electric current should be the largest and the resistance is the smallest. Fig. 5(d) shows the angular dependence of the $\left(\sigma_{XX}(\theta)-\sigma_{XX}(0)\right)$ in the Mn$_{1.25}$Ga/Pt and Mn$_{1.25}$Ga/Pt bilayers. The applied current is along X axis, and the magnetic field of 9 T is applied in the ZX plane with angle $\theta$ relative to the Z axis. We have observed the decrease of $\left(\sigma_{XX}(\theta)-\sigma_{XX}(0)\right)$ in the Mn$_{1.25}$Ga/Al bilayers as increasing $\theta$, reflecting the anisotropic magnetoresistance of the single Mn$_{1.25}$Ga layer as shown in our previous work [36]. However, the value in the Mn$_{1.25}$Ga/Pt bilayers have been enhanced with increasing $\theta$, which is consistent with the assumption that there is an additional electric current in the Mn$_{1.25}$Ga/Pt bilayers and it is the largest when $\mathbf{E}/\!/\mathbf{B}$. Therefore, these transport features may be related to the Weyl fermions at the Mn$_x$Ga/Pt interfaces. Recent theory and experiments have suggested that the Weyl fermions with broken time-reversal symmetry were also expected to generate strong intrinsic AHE [37, 38], which may also contribute to the UAHE in the low temperature limit. However, the transport properties of our bilayer systems are more complicated since bulk Pt, bulk $Mn_xGa$ and $Mn_xGa$/Pt interfaces will all contribute to the transport features [18]. Other mechanisms including orbital magnetoresistance are hard to be distinguished [39]. It would be interesting to verify the possibility of interface-driven Weyl fermions in future studies.

In conclusion, we have discussed the origin of interface-driven UAHE in the $Mn_xGa$/Pt bilayers, which was distinguished from the THE related to chiral spin structures. Low temperature MFM results indicate that the additional Hall signals as a function of magnetic field does not peak near the maximal density of magnetic bubbles. We firmly demonstrates that the bump/dip features of Hall signals in FM/HM heterostructures cannot be taken as an unambiguous signature for chiral spin structures, and it should be treated discreetly since a wealth of underlying and interesting physics are often regrettably missed. It further demonstrates that an extraordinary range of unanticipated phenomena is expected to emerge from the interfaces, exciting new scientific results and functionality for future research.

**Acknowledgements**

This work was partially supported by the National Science Foundation of China (Grant Nos. 51731003, 51671019, 51602022, 61674013, 51602025, 11675023), and the Fundamental Research Funds for the Central Universities (FRF-BD-18-014A).

**Figure captions**

**Fig. 1.** (a)-(c) The Hall resistivities after subtracting ordinary Hall contributions $\left(\rho_{XY} - \rho_O\right)$ of the Mn$_x$Ga/Pt bilayers under perpendicularly applied magnetic fields in the temperature range from 5 to 300 K; (d)-(f) MR of the Mn$_x$Ga/Pt bilayers under perpendicularly applied magnetic fields in the temperature range from 5 to 300 K.

**Fig. 2.** (a)-(c) $\rho_A$ *vs* $\rho_{xx}^2$ for $Mn_xGa/Pt$ bilayers. Solid red lines refer to a linear fit of $\rho_A = \alpha\rho_{XX0} + b\rho_{XX}^2$; (d) Temperature dependence of skew scattering coefficients $\alpha$ for $Mn_xGa/Pt$ bilayers.

**Fig. 3.** (a)-(c) The additional Hall resistivities after subtracting ordinary and anomalous Hall contributions $\Delta\rho_{XY}$ of the $Mn_xGa/Pt$ bilayers as functions of perpendicularly applied magnetic fields at different tempratures from 5 to 300 K. $\Delta\rho_{XY}^{max}$ denotes the peak (300 K) or valley (5 K) values with sweeping magnetic field from positive to negative; (d) Temperature dependence of $\Delta\rho_{XY}^{max}$ for the $Mn_xGa/Pt$ bilayers; Magnetic force microscopy images for the $Mn_{1.55}Ga/Al$ (e) and $Mn_{1.55}Ga/Pt$ bilayers (f) at 300 K after applying a positive perpendicular magnetic field of 5 T and then decreasing the magnetic field down to zero. The scale size is 7 μm×5 μm.

**Fig. 4.** $M$-$H$ (a) and $\Delta\rho_{XY} - H$ (b) curves for $Mn_{1.55}Ga/Pt$ bilayers at 3.5 K. (c)-(h) Magnetic force microscopy images for the after applying a positive perpendicular magnetic field of 5 T and then decreasing the magnetic field down to +2 T, 0 T, -0.5 T, -1.0 T and -2.0 T, respectively. The scale size is 7 μm×7 μm.

**Fig. 5.** (a) Mn $2p$ XAS spectra of the $Mn_{1.25}Ga/Pt$ and $Mn_{1.25}Ga/Al$ bilayers. (b) Temperature dependence of $\Delta\sigma_{XY}^{max}$ for the $Mn_xGa/Pt$ bilayers; (c) Temperature dependence of $d^2\Delta\sigma_{XY}^{max}/dT^2$ and $d^2\alpha/dT^2$ for the $Mn_{1.55}Ga/Pt$ bilayers; (d) Angle dependence of $\left(\sigma_{XX}(\theta) - \sigma_{XX}(0)\right)$ at 5 K for $Mn_{1.25}Ga/Al$ and $Mn_{1.25}Ga/Pt$ bilayers. The current is applied along X direction, and θ=0 ° denote the magnetic field (9 T) is applied along Z direction. The signals in $Mn_{1.25}Ga/Pt$ bilayers are multiplied by a factor of 10 shown as ×10.

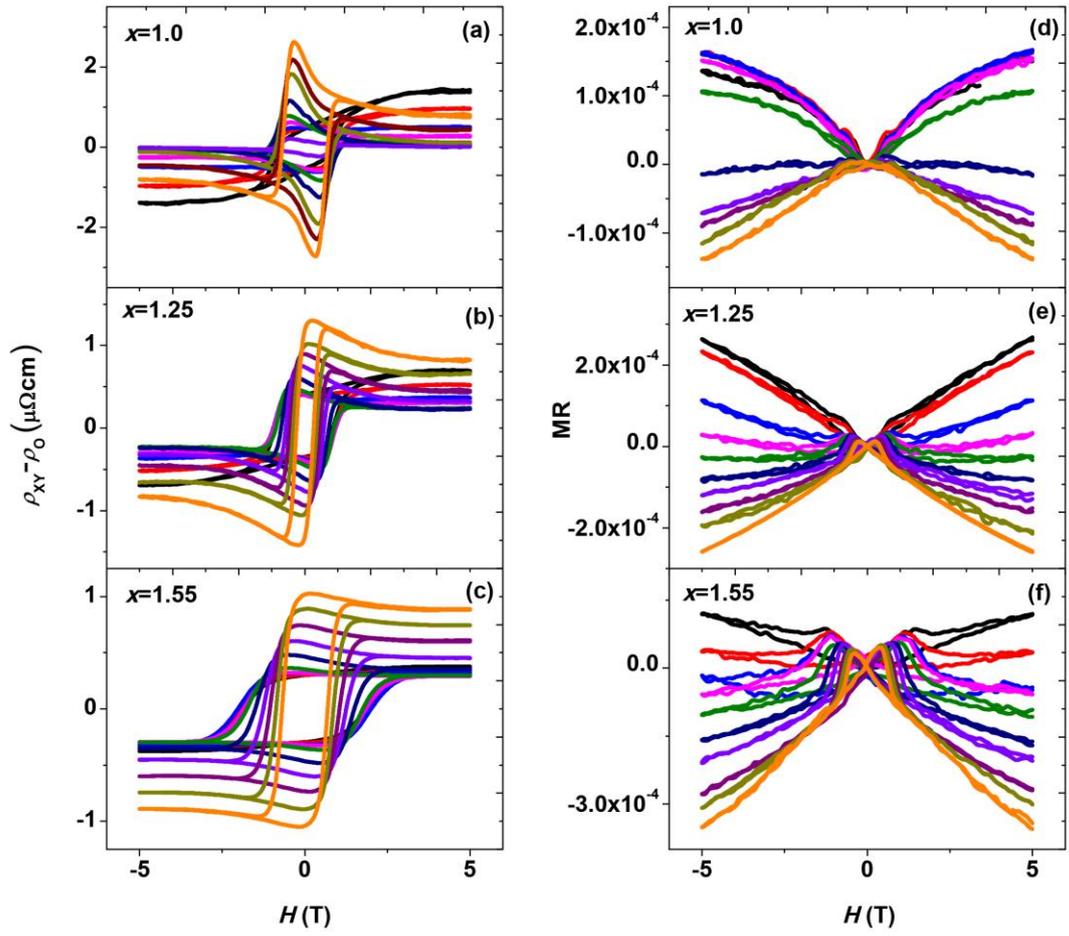

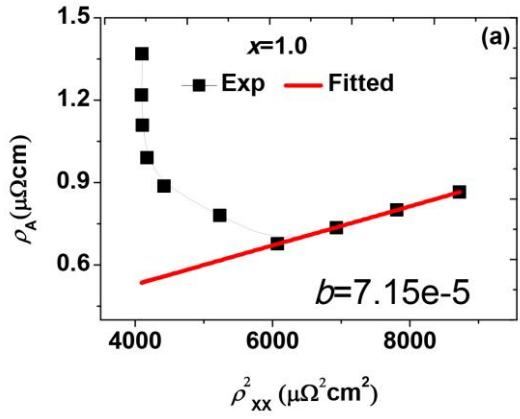

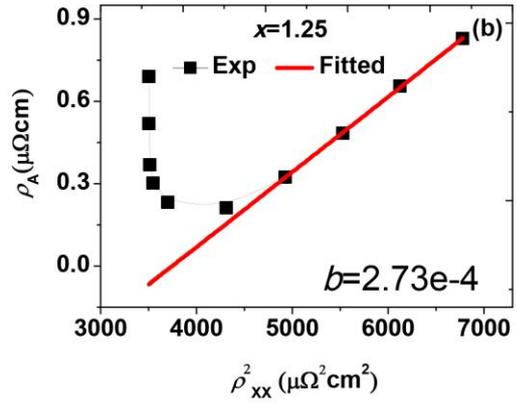

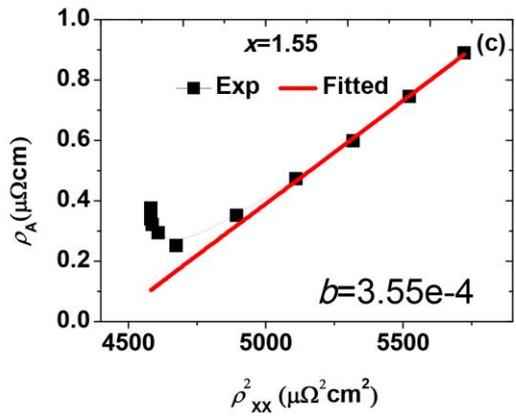

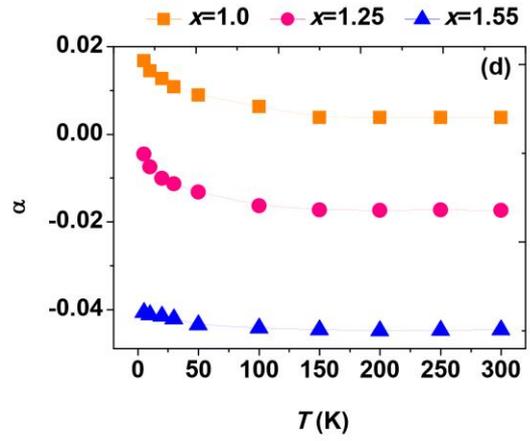

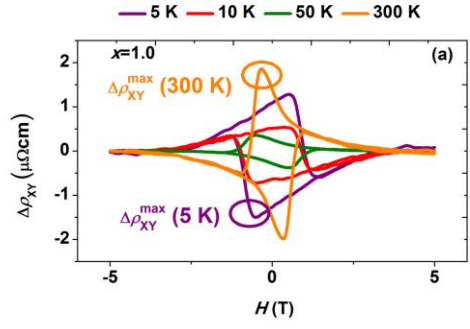

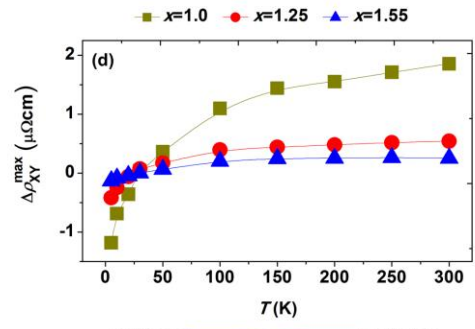

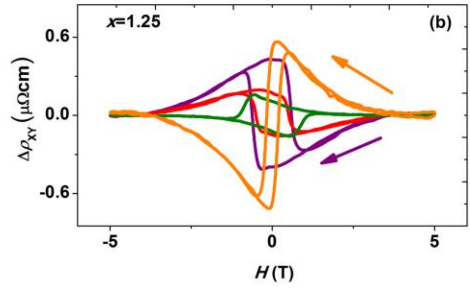

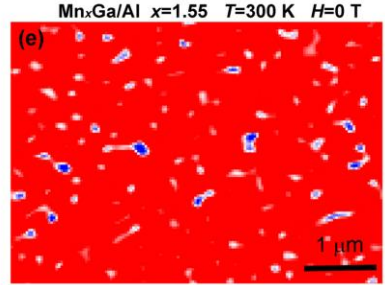

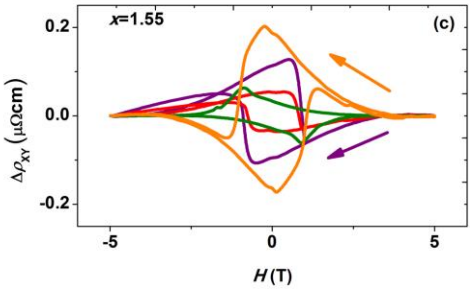

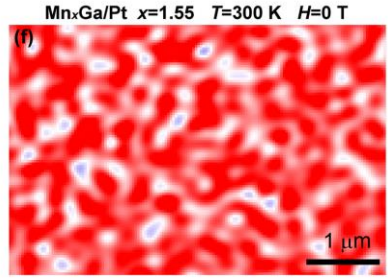

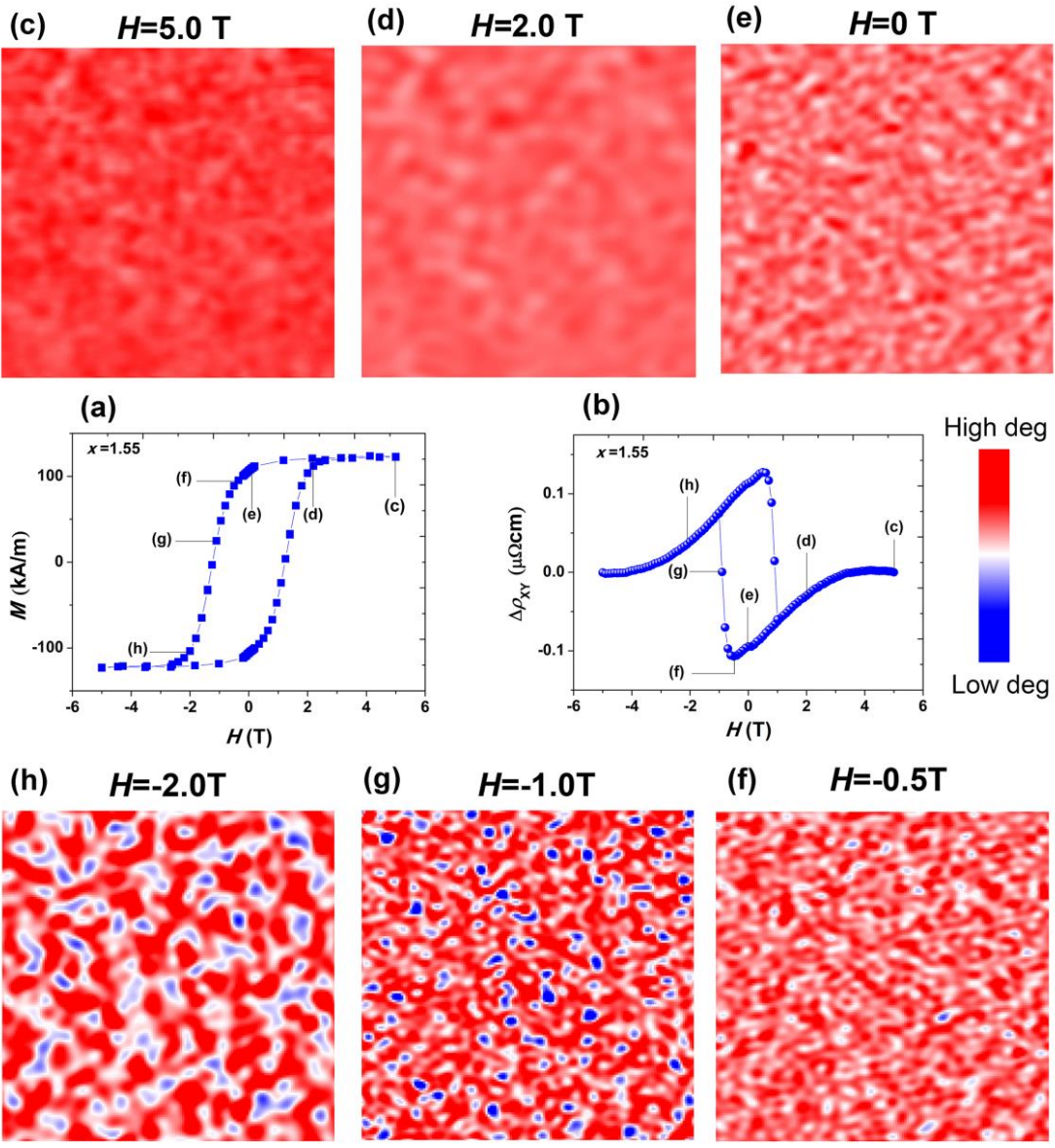

(c) *H*=5.0 T  (d) *H*=2.0 T  (e) *H*=0 T

(a)

(b)

High deg

Low deg

(h) *H*=-2.0T  (g) *H*=-1.0T  (f) *H*=-0.5T

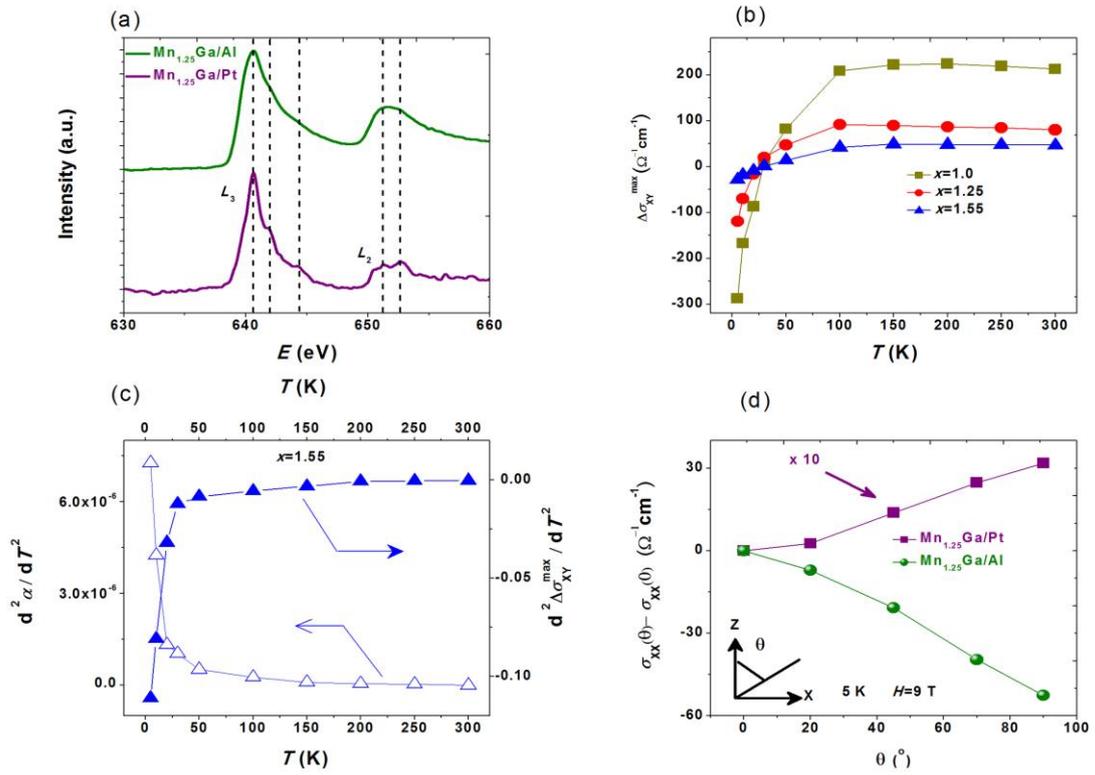